# Detection of boson peak and fractal dynamics of disordered system using terahertz spectroscopy

Tatsuya Mori[a,1], Yue Jiang[a], Yasuhiro Fujii[b], Suguru Kitani[c], Hideyuki Mizuno[d], Akitoshi Koreeda[b], Leona Motoji[a], Hiroko Tokoro[a], Kentaro Shiraki[a], Yohei Yamamoto[a], Seiji Kojima[a]

[a]Division of Materials Science, University of Tsukuba, 1-1-1 Tennodai, Tsukuba, Ibaraki 305-8573, Japan

[b]Department of Physical Sciences, Ritsumeikan University, 1-1-1 Nojihigashi, Kusatsu 525-8577, Japan

[c]Laboratory for Materials and Structures, Tokyo Institute of Technology, 4259 Nagatsuta-cho, Midori-ku, Yokohama 226-8503, Japan

[d]Graduate School of Arts and Sciences, The University of Tokyo, Tokyo 153-8902, Japan

## Abstract

Disordered systems exhibit universal excitation, referred to as the boson peak, in the terahertz region. Meanwhile, the so-called fracton is expected to appear in the nanoscale region owing to the self-similar structure of monomers in polymeric glasses. We demonstrate that such excitations can be detected using terahertz spectroscopy. For the interaction between terahertz light and the vibrational density of states of the fractal structure, we formulate an infrared light-vibration coupling coefficient for the fracton region. Accordingly, we show that information concerning fractal and fracton dimensions appears in the exponent of the absorption coefficient. Finally, using terahertz time-domain spectroscopy and low-frequency Raman scattering, we experimentally observe these universal excitations in a protein lysozyme system that has an intrinsically disordered and self-similar nature in a single supramolecule. These findings are applicable to disordered and polymeric glasses in general and will be key to understanding universal dynamics of disordered systems by terahertz light.

Keywords: boson peak, fractons, terahertz spectroscopy, lysozyme

## Introduction

In disordered materials, universal excitation, i.e., the so-called boson peak (BP), appears in the terahertz (THz) region, where propagating sound waves end, as an anomaly in the acoustic phonon mode[1]. The Debye theory predicts the behaviour of the acoustic phonon of a crystalline system, where the vibrational density of states (v-DOS, $g(\nu)$) follows the power law, $g(\nu) \propto \nu^{D-1}$, in a $D$-dimensional system. The BP indicates a deviation from the Debye theory, and the peak appears universally in the spectrum of $g(\nu)/\nu^{D-1}$ as an excess of the v-DOS in amorphous solids. While closely related to low-temperature universal thermal properties in glassy materials[2,3], the universal excitation has long been an unresolved issue in glass physics.

Meanwhile, in a polymer glass system, the fractal structure of a polymer chain, which is induced by the self-similar connectivity of structural units, will lead to fractal dynamics[4,5]. The dynamics of fractal objects, referred to as *fracton,* follow $g(\nu) \propto \nu^{d_f-1}$, where $d_f$ is the fracton dimension that typically takes the value of 4/3 and does not depend on the Euclidean dimension $D$, as conjectured by Alexander and Orbach[4].

Interestingly, such universal dynamics of disordered materials also appears in a few single crystals. As discovered in recent studies, BPs or amorphous-like thermal behaviours appear in single crystals, such as thermoelectric materials[6–9], pure relaxor materials[10], and atypical organic materials[11,12], and theoretical studies for elucidating anomalous phenomena are underway[8,13–15]. In these materials, the off-centre large-amplitude anharmonic vibration modes play a crucial role in the universal glass-like thermal properties of the crystals. Furthermore, the fracton in a single crystal appears in pure relaxor substances[16,17] owing to the connection of dipole-dipole moments having a fractal structure, which might be key to explaining the large dielectric response.

With regard to actual glassy materials, the most direct and well-known methods for detecting BP excitation are inelastic neutron or X-ray scattering (INS/IXS) and inelastic nuclear scattering, those are used to directly investigate the v-DOS[18–21]. Whereas numerous studies have investigated BPs with low-frequency Raman scattering[3,22–24] and low-temperature specific heat[2,3,25,26], few recent THz spectroscopic studies on BP dynamics have been reported[27–30]. In addition, theoretical[31–35] and molecular dynamics (MD) simulation[36–39] studies on BPs are rapidly advancing. However, relatively few studies have focused on fractons[5,40–42], particularly on nanoscale dynamics. Moreover, there are practically no studies of fractons by THz spectroscopy[43] owing to insufficient understanding of the interaction between THz light and v-DOS in the fracton region, even though numerous THz spectroscopic studies have investigated polymer substances in the last decade[44,45].

A protein lysozyme employed in this study is an example of a polymer glass-like substance, used to investigate the universal dynamics of BP and fractons using THz spectroscopy. Although a protein molecule is not an actual glassy material, low-frequency Raman scattering has shown[46] that its BP dynamics appear similar to those of glassy materials, even in the case of a single crystal of protein. This implies that the protein system has an intrinsically disordered and polymeric nature within a single supramolecule, where each amino acid corresponds to a monomer molecule of polymer glass. The structural information of a single crystal of protein from the Protein Data Bank (PDB) makes it possible to discuss the correlation of the structure and its dynamics.

In this work, we first formalise the infrared (IR) light-vibration coupling coefficient $C_{IR}(v)$ in the fracton region, where we modify the $C_{IR}(v)$ model proposed by Taraskin *et al.*[47] We find that the exponent of $C_{IR}(v)$ is almost identical to that of the Raman coupling coefficient $C_{Raman}(v)$. Second, in the THz spectrum of lysozyme, we observe the BP and fractons, and find that the slope of the absorption coefficient in the log-log plot is identical to that of the Raman spectrum. Third, we find that the observed value of the slope in the fracton region is in good agreement with the value predicted by the formalised $C_{IR}(v)$ and $C_{Raman}(v)$. Furthermore, we find that the slope is characterised by fractal and fracton dimensions. Finally, we estimate the structure correlation length associated with the BP and the characteristic wavelengths of the IR and Raman active modes. The results suggest that these universal dynamics occur within a single protein molecule.

## Results
### Formalism of $C_{IR}$ in the fracton region.

First, we briefly introduce the dispersion relation of a fractal system[4,42,48]. The mean square displacement $\langle R^2 \rangle$ of a fractal object behaves as an anomalous diffusive process:

$$\langle R^2 \rangle \propto t^{2\xi}, \quad (1)$$

where $\xi < 1/2$; this anomalous diffusive process is more localised than a normal diffusive process ($\xi = 1/2$). In this case, the autocorrelation function becomes $P_0(t) \propto t^{-D_f \xi}$, where $D_f$ is the fractal dimension, and we immediately obtain v-DOS, as follows:

$$g(v) \propto v^{d_f - 1}, \quad (2)$$

where $d_f$ is the fracton dimension. Vibrational modes were originally referred to as 'fractons' by Alexander and Orbach[4]. The relation among $\xi$, $D_f$, and $d_f$ is $\xi = d_f / 2D_f$, and the anomalous diffusive process with $\xi < 1/2$ leads to $d_f < D_f$. Therefore, for a fractal object, the dynamic fracton dimension is generally smaller than the static fractal dimension. In a study by Rammal and Toulouse[48], the dispersion relation of a fractal object with the assumption of length scaling was expressed as:

$$v \propto k^{\frac{D_f}{d_f}}, \quad (3)$$

where $k$ is a wave number. In this work, we use this dispersion relation for the fracton region of a protein system. Note that $2\pi k^{-1}$ represents the localised wavelength of the fracton mode rather than the wavelength of a propagating plane wave.

Next, we show how $C_{IR}(\nu)$ behaves in a fractal system by applying the dispersion relation to a universal model proposed by Taraskin *et al.*[47] According to the model, the starting point for $C_{IR}(\nu)$ is expressed as follows[49,50]:

$$C_{IR}(\nu) = \frac{2\pi^2 n}{c\sqrt{\varepsilon_\infty}} \left| \sum_i \frac{q_i}{\sqrt{m_i}} \mathbf{e}_i(\nu) \right|^2, \tag{4}$$

where $q_i$, $m_i$, and $\mathbf{e}_i(\nu)$ are the fixed but spatially fluctuating atomic charges, mass, and vibrational eigenvector of frequency $\nu$ corresponding to atom $i$, respectively. Further, $\varepsilon_\infty$ represents the high-frequency dielectric constant and $n$ is the atomic concentration. This expression is derived from the linear response theory in the harmonic approximation for atomic vibrations[49]. In the theory of Taraskin *et al.*, $q_i$ is divided into two parts: one is the uncorrelated charge, $q_{1i}$, and the other is the correlated charge, $q_{2i}$. In their interpretation, $q_{2i}$ satisfies local charge neutrality whereas $q_{1i}$ indicates the randomly fluctuating charges that deviate from the neutral charge, i.e., in the crystal system, $q_{1i}$ becomes 0 and $q_{2i} = q_i$. The charge components can then be recast in terms of uncorrelated and correlated charge components as $S_1$ and $S_2$, respectively:

$$C_{IR}(\nu) = \left\langle \frac{2\pi^2 n}{c\sqrt{\varepsilon_\infty \bar{m}}} \left| \sum_i (q_{1i} + q_{2i})\, e^{i\mathbf{k}\cdot\mathbf{r}_i} \right|^2 \right\rangle = \frac{2\pi^2 n}{c\sqrt{\varepsilon_\infty \bar{m}}} \langle |S_1 + S_2|^2 \rangle \tag{5}$$

If there is no correlation between $q_{1i}$ and $q_{2i}$, the above-mentioned formula reduces to:

$$C_{IR}(\nu) = \frac{2\pi^2 n}{c\sqrt{\varepsilon_\infty \bar{m}}} (\langle |S_1|^2 \rangle + \langle |S_2|^2 \rangle). \tag{6}$$

It is easily found that the first term is a constant[47,50]. Regarding the second term, $q_{2i} = \sum_{j\neq i} \Delta q_{ji}$, i.e., $q_{2i}$ can be considered to result from charge transfers between nearest neighbours, where $j$ runs through all the nearest neighbours of atom $i$, $\Delta q_{ji} = -\Delta q_{ij}$ represents the charge transfer from the originally neutral atom $j$ to the originally neutral atom $i$, and $\mathbf{k}\cdot\mathbf{r}_{ij} \ll 1$, considering the long wavelength regime[50]. The component of $S_2$ is approximated by the right-hand side of the following equation:

$$S_2 = \left\langle \sum_{(ij)} \Delta q_{ij}(e^{i\mathbf{k}\cdot\mathbf{r}_j} - e^{i\mathbf{k}\cdot\mathbf{r}_i}) \right\rangle \cong k \sum_{(ij)} \Delta q_{ij} e^{i\mathbf{k}\cdot\mathbf{r}_{(ij)}} (i\hat{\mathbf{n}} \cdot \mathbf{r}_{ij}) \tag{7}$$

where $\mathbf{r}_{ij} = \mathbf{r}_j - \mathbf{r}_i$, $\mathbf{r}_{(ij)} = (\mathbf{r}_j + \mathbf{r}_i)/2$ and $\mathbf{k} = k\hat{\mathbf{n}}$ are defined. We therefore see that the component of $S_2$ is proportional to the value of $k$. When the dispersion relation of the Debye model, i.e., $\nu \propto k$, is substituted for the above-mentioned relation, we obtain the quadratic frequency dependence of the second term of $C_{IR}(\nu)$:

$$\langle |S_2|^2 \rangle \propto k^2 \propto \nu^2. \quad (8)$$

Then, Taraskin's universal functional form of $C_{IR}(\nu)$ is expressed as follows:

$$C_{IR}(\nu) = A + B\nu^2. \quad (9)$$

Here, we consider the frequency region above the BP frequency[41], where we expect the appearance of *fracton modes*. We use the dispersion relation of fracton modes, Eq. (3), and substitute it, instead of the Debye model, for Eq. (7). The second term of $C_{IR}(\nu)$ then becomes:

$$\langle |S_2|^2 \rangle \propto k^2 \propto \nu^{2\frac{d_f}{D_f}}, \quad (10)$$

where the frequency exponent is changed from 2 to $2d_f/D_f$. We thus obtain the expression of $C_{IR}(\nu)$ in the fracton region as:

$$C_{IR}(\nu) = A + B\nu^{2\frac{d_f}{D_f}}. \quad (11)$$

Considering the behaviour of the v-DOS expressed in Eq. (2), we expect the frequency dependence of absorption coefficient $\alpha(\nu)$ to be proportional to $\nu^{2d_f/D_f+d_f-1}$.

Note that, according to a study by Boukenter *et al.*[51]:

$$C_{Raman}(\nu) = \nu^{2\frac{d_f d_\phi}{D_f}}, \quad (12)$$

and we find that the second term of $C_{IR}(\nu)$ is identical to $C_{Raman}(\nu)$, except for the super-localized exponent $d_\phi$; in many cases, $d_\phi$ becomes 1. We can therefore expect the THz and Raman spectra to show the same gradient in the fracton region.

**Boson peak of lysozyme in THz and Raman spectra.**

Figs. 1a and 1b show the real part $\varepsilon'(\nu)$ and the imaginary part $\varepsilon''(\nu)$, respectively, of the complex dielectric constant of lysozyme at room temperature. The spectrum of $\varepsilon''(\nu)$ shows a broad absorption band, which is typical of amorphous materials, where a convex upward curve appears > 0.6 THz and a broad peak appears at 3.26 THz. Correspondingly, $\varepsilon'(\nu)$ shows Debye relaxation mode-like behaviour[30]. However, if we observe the detailed structure around 0.6 THz, we find that the real part has a small peak, indicating that the spectrum below 0.6 THz shows more resonant-like behaviour rather than Debye-like behaviour. These features are prominent at lower temperatures, as shown in the insets of Figs. 1a and b, and similar behaviour is generally observed in glassy materials (see Fig. 1 of ref. [30]). Such resonant-like deviation, suggesting the existence of the BP at around 0.6 THz, is a universal behaviour of the dielectric constant, owing to the characteristic spectral shape of the v-DOS around the BP frequency and the breakdown of selection rules caused by the disappearance of translational symmetry, as shown in a previous study[30].

To investigate the BP of lysozyme in the IR spectrum, the spectrum of the complex dielectric constant is converted into $\alpha(v)/v^2$ (see Table S1), i.e., the BP plot of the IR spectrum, with the result shown in Fig. 1c. According to the $\alpha(v)/v^2$ spectrum, the BP of lysozyme is clearly observed at 0.6 THz. We can therefore conclude that detection of the BP using THz spectroscopy is possible for proteins, as well as for actual glassy materials.

The BP dynamics of lysozyme have also been investigated using low-frequency Raman scattering. Fig. 1d shows the measured Raman intensity $I(v)$ at room temperature, and a clear peak is observed at around 0.8 THz. In general, the Raman susceptibility $\chi''(v)$ is extracted from the measured Raman intensity $I(v)$ through the following equation[52,53]:

$$I(v) = [n_B(v,T) + 1] \cdot \chi''(v), \tag{13}$$

where $n_B(v,T) = (\exp(hv/k_BT) - 1)^{-1}$ is the Bose–Einstein distribution function. The obtained $\chi''(v)$ spectrum is shown in Fig. 1e and shows a broad peak at 2.53 THz. We observe that this peak frequency is lower than that of the infrared $\varepsilon''(v)$ spectrum (3.26 THz). The BP plot of the Raman spectra is $\chi''(v)/v$, and it is often referred to as reduced intensity. The relation between $\chi''(v)$ and $g(v)$ is typically expressed by the following equation[54]:

$$v \cdot \chi''(v) = C_{\text{Raman}}(v) \cdot g(v). \tag{14}$$

For convenience, the BP plot relations are summarised in Table S1 in supplementary information.

As shown in Fig. 1f, in the $\chi''(v)/v$ spectrum, the BP of the reduced Raman intensity spectrum is observed at around 0.8 THz, which is slightly higher than the BP frequency of the IR spectrum $v_{\text{BP-IR}}$, as discussed below. We emphasise that no absorption peak is observed in the $\chi''(v)$ spectrum near the BP frequency of the reduced Raman intensity $v_{\text{BP-Raman}}$, as well as the $\varepsilon''(v)$ of the IR spectrum[30]. However, as shown in Fig. 1d, we clearly observe the BP in the measured $I(v)$ at room temperature, as with other glassy materials. Even though its direct observation at room temperature has historically made Raman spectroscopy a well-known method for detecting the BP, we point out that the 'boson peak' in the measured $I(v)$ at room temperature is not a direct peak derived from a particular mode but a 'universal artefact', as discussed in our previous work[30]. The appearance of the BP in $I(v)$ is understood by high-temperature approximation, where the thermal factor $n_B(v,T) + 1$ of Eq. (13) becomes $1/v$, because the energy at room temperature, $k_BT$, is approximately six times higher than the typical energy of the BP frequency of 1 THz (cf. 1 THz = 33.3 cm$^{-1}$ = 4.14 meV = 48 K). The approximation results on the right-hand side of Eq. (13) transform into $\chi''(v)/v$, i.e., the measured $I(v)$ at room temperature becomes identical to the BP plot of the Raman spectra.

**Comparison between $\varepsilon''(v)$, $\chi''(v)$, and $g(v)$ and determination of coupling coefficients.**

We directly compare IR, Raman, and $g(v)$ spectra, and determine the light-vibration coupling coefficients, $C_{IR}(v)$ and $C_{Raman}(v)$. Figs. 2a and 2b show the imaginary parts of $\varepsilon''(v)$ and $\chi''(v)$ and the BP plots of the IR and Raman spectra on the same axis, respectively. From the comparison of the BP plots shown in Fig. 2b, we find that $v_{BP-Raman}$ is 1.4 times higher than $v_{BP-IR}$, and the difference is phenomenologically attributed to the different frequency dependences of $C_{IR}(v)$ and $C_{Raman}(v)$. Meanwhile, there is a perfectly overlapped region of $\varepsilon''(v)$ and $\chi''(v)$ above the BP frequency, where the convex upward line shapes show the same exponential behaviour as expected in the fracton region. To ascertain the exponential behaviour, we create a log-log plot of $\varepsilon''(v)$ and $\chi''(v)$, as shown in Fig. 3a. The same linear slope region appears from around 0.8 THz up to 2.2 THz, and constitutes the fracton region.

To evaluate the interaction between light and v-DOS ($g(v)$) in the BP and fracton regions, we determine $C_{IR}(v)$ and $C_{Raman}(v)$ by combining $g(v)$ with our data of $\varepsilon''(v)$ and $\chi''(v)$. The $g(v)$ data are obtained from the results of INS measurements[40,55], and the depicted $g(v)$ and BP plots $g(v)/v^2$ are shown in Figs. 3a and 2c, respectively. Furthermore, to obtain the absolute value of $C_{IR}(v)$, we determine the absolute value of $g(v)$ by fitting $g_{heat}(v)$, which is extracted from the low-temperature specific heat data[25] through maximum entropy method (MEM) analysis (for further details, see Fig. S1). In the $g(v)/v^2$ spectrum shown in Fig. 2c, we find that the BP frequency $v_{BP-INS}$ is 0.58 THz, which is obtained via log-normal function fitting[56]. In the log-log plot of $g(v)$ shown in Fig. 3a, $v_{BP-INS}$ becomes a boundary where the upper and lower regions show different gradients of power-law dependence. Below $v_{BP-INS}$, $g(v)$ becomes proportional to $v^{D-1}$, where $D$ is 3, indicating that the v-DOS precisely follows the Debye model of the $3D$ system. In contrast, above $v_{BP-INS}$, the value of the exponent decreases to 0.4, which is expected in the fracton region where the exponent is expressed as $d_f - 1$[4]. Then, $C_{IR}(v)$ with the absolute value and $C_{Raman}(v)$ of lysozyme are experimentally determined through the following relation[57]:

$$\alpha(v) = C_{IR}(v) \cdot g(v), \qquad (15)$$

and Eq. (14). The results are shown in Fig. 3b and its inset. The related equations are summarised in Table S1.

**Analysis of coupling coefficients: Boson peak.**

The $C_{IR}(v)$ in the vicinity of the BP frequency is evaluated quantitatively using Taraskin's model[47]. Further, $C_{Raman}(v)$ is explained by the coupling of light with elastic strains via spatially fluctuating elasto-optic constants[58]. In the following, we focus on analysis of $C_{IR}(v)$.

First, we focus on the behaviour of $C_{IR}(v)$ and $C_{Raman}(v)$ around the BP frequency, as shown in the inset of Fig. 3b. In the vicinity of the BP frequency, the $C_{IR}(v)$ of lysozyme consists of a constant term and a quadratic or higher-order term, and the behaviour is similar to the $C_{IR}(v)$ of silica glass[47,59], i.e., the $C_{IR}(v)$ is dominated by a constant term below the BP frequency. In contrast, the $C_{Raman}(v)$ of lysozyme shows almost linear frequency dependence, which is common for numerous glassy substances[24]. As a result, $v_{BP-IR}$ is almost the same as $v_{BP-INS}$, while $v_{BP-Raman}$ becomes slightly higher than $v_{BP-INS}$.

In the following, we analyse the frequency dependence and absolute value of $C_{IR}(v)$. For the $C_{IR}(v)$ of glassy materials, Taraskin *et al*. derived a universal functional form[47,50] by applying the linear response theory to the disordered system. The result is given by Eq. (9), where the constant term $A$ originates from the response of the uncorrelated charge denoted by $q_1$, i.e., the randomly fluctuating charges that deviate from the average charge of each atom. Further, the higher-order term $B$ is related to $q_2$, which denotes the average charge value, and this term plays a dominant role above the BP frequency. We discuss the origin of the absorption coefficient of lysozyme by analysing $C_{IR}(v)$ on the basis of Taraskin's model. We obtain the value of $A$, which is dominant in the vicinity of the BP frequency, and the result is 1399 cm$^{-1}$THz. We then use the following relation of Taraskin's model to evaluate the uncorrelated charge $q_1$:

$$A = \frac{\langle q_{1i}^2 \rangle 2\pi^2 n}{\bar{m} c \sqrt{\varepsilon_\infty}}, \tag{16}$$

where $n = \rho/\bar{m}$, $\bar{m} = N^{-1} \sum_i m_i$ ($i$ is the number of atoms), $N$ is the number of atoms in a solid of volume $V$, and $\rho$ is the density. Then, we extract the value of the $q_1$ of lysozyme as 0.127 $e$, with other parameters listed in Table S2. At 1 THz, the absorption coefficient of lysozyme is 21 times greater than that of silica glass[47,59], which is a typical network glass former (see Table S2). Regarding the absolute value of $C_{IR}(v)$, the $A$ of lysozyme is 16 times greater than that of silica glass[47,59]. This result implies that the large $C_{IR}(v)$ is responsible for lysozyme's large absorption in the THz region. We further consider the origin of this large $C_{IR}(v)$. As compared to silica glass, the uncorrelated charge $q_1$ of lysozyme is 2.12 times larger; the contribution of the uncorrelated charge term $\langle q_{1i}^2 \rangle$ is thus 4.48 times larger than that of silica glass. In addition, the average mass, $\bar{m}$, of lysozyme is 0.38 times smaller than that of silica glass. As a result, the affective term, $1/\bar{m}^2$, becomes 6.93 times larger than that of silica glass, thus making a larger contribution compared to the uncorrelated charge. We therefore point out that an organic glass former with a relatively low mass generally shows a larger constant term $A$ in the vicinity of the BP frequency compared to an inorganic material, owing to the effect of its lower mass, independent of the effect of the uncorrelated charge.

**Analysis of coupling coefficients: fracton.**

Next, we shed light on the fracton behaviour that appears above the BP frequency, where both $\varepsilon''(v)$ and $\chi''(v)$ exhibit a power-law behaviour, as shown in Fig. 3a. This is because both the v-DOS and coupling coefficients show a power-law behaviour.

As shown in Fig. 3a, $\varepsilon''(v)$ and $\chi''(v)$ are plotted in the log-log representation and they show linear frequency dependence above the BP frequency. Interestingly, and as expected, the IR and Raman spectra share a frequency region with almost identical slopes, and the region extends up to the cut-off of the Raman spectrum at 2.53 THz. On the contrary, the linearity of the IR spectrum continues towards higher frequencies and is cut off at 3.26 THz. In addition, the v-DOS $(g(v))$[40,55] is simultaneously plotted to show the fracton region directly. Note that $g(v)$ shows linearity of the fracton above 1.7 THz, although the starting frequency of the fracton is slightly higher than that of IR and Raman spectra. Such power-law behaviours of both IR and Raman spectra and of the v-DOS spectrum generate linear regions in both $C_{IR}(v)$ and $C_{Raman}(v)$. These linear regions are the fracton regions, and the exponent values for $C_{IR}(v)$ and $C_{Raman}(v)$ are obtained as 1.10 and 1.00, respectively (see the left column of Table 1), from the fitting in Fig. 3b. These exponent values are what we expect to be represented by $2d_f/D_f$ in this study.

Here, we examine whether the performed formalism of $C_{IR}(v)$ is suitable for expressing observed behaviour by evaluating the exponent value of $C_{IR}(v)$ with the individually obtained $D_f$ and $d_f$ from theoretical calculations and experiments.

First, we verify the $C_{IR}(v)$ formula with the calculated mass fractal dimension $D_{f\text{-calc}}$ and fracton dimension $d_{f\text{-calc}}$. We calculate the mass fractal dimension of lysozyme as $D_{f\text{-calc}} = 2.75$ using the structural information of 1LYZ obtained from the PDB, as shown in Figs. 4a and 4b. $D_{f\text{-calc}}$ is defined as the slope obtained when plotting the mass of all atoms contained in concentric spheres of radius $R$ from the centre of gravity of the lysozyme molecule on a log-log scale[60]. The obtained value of $D_{f\text{-calc}}$ is consistent with that reported previously[60]. The value of $d_{f\text{-calc}}$ is taken from literature, where the harmonic spectrum of the vibrational modes calculated by a Gaussian network model determines the fracton dimension as $d_{f\text{-calc}} = 1.43$ for approximately 100 amino acids[61]. By substituting $D_{f\text{-calc}}$ and $d_{f\text{-calc}}$ into the $C_{IR}(v)$ formula, the calculated slope of $C_{IR}(v)$ is $2d_{f\text{-calc}}/D_{f\text{-calc}} = 1.04$, in excellent agreement with our experimental value of 1.10 (see Table 1).

Second, we verify the $C_{IR}(v)$ formalism using the values of $D_{f\text{-exp}}$ and $d_{f\text{-exp}}$, which are experimentally determined utilizing small angle neutron scattering (SANS) and INS[40], respectively. The fractal dimension, $D_{f\text{-exp}} = 2.78$, of lysozyme is obtained by the power-law dependence of SANS intensity on the wavenumber in the double logarithmic scale, within a spatial region of 2.6–4.4 nm, suggesting that the mass fractal of lysozyme appears within a single lysozyme molecule with a diameter of approximately 3.7 nm[62]. Furthermore, the fracton dimension, $d_{f\text{-exp}} = 1.43$, of lysozyme is obtained from the v-DOS determined by the INS experiment[40]. A linear property is observed and can be divided into two slopes of 1.99 and 0.43, below and above the BP frequency, respectively, with the BP as the boundary. The former is in good agreement with the behaviour of the 3*D* Debye model. The latter corresponds to the value of $d_f - 1$, which is the fracton behaviour of the v-DOS. By substituting the experimentally determined $D_f$ and $d_f$, the slope of $C_{IR}(v)$ is calculated as $2d_f/D_f = 1.02$. This result is also in good agreement with our experimentally determined slope of 1.10. The above-mentioned examinations thus confirm the validity of the proposed $C_{IR}(v)$ in the fracton region.

We verified the $C_{IR}(v)$ formalism proposed in this study through theoretical calculations and experimental results. We can thus quantitatively describe the slope of the absorption coefficient $\alpha(v)$ observed in experimental data using the $C_{IR}(v)$ formalism of the fracton region. The results show that the concept of the fracton is adequate for expressing the observed power-law behaviour of the region above the BP frequency of the protein lysozyme in the IR and Raman spectra. Even though we conducted tests only on the protein lysozyme, our results are applicable not only to other proteins but also universally to other polymeric glasses with a fractal structure.

**Hypothesis about continuity of $C_{IR}(v)$ and characteristic lengths.**

Finally, through an additional step considering the dispersion relation used for the coupling coefficient, we discuss the end mode of the fracton region. The expression of Eq. (15) is approximated from $\alpha(v) = \sum C_i(v) g_i(v) = C_{LA}(v) g_{LA}(v) + C_{TA}(v) g_{TA}(v) + other\ terms^{57}$. Generally, the transverse acoustic (TA) mode plays a dominant role in the vicinity of the BP frequency owing to the smaller slope of the dispersion relation compared to the longitudinal acoustic (LA) mode, resulting in larger v-DOS of the TA mode. In addition, based on the fact that neither $\alpha(v)$ nor $g(v)$ show any jumps in the vicinity of the BP in our system, we create the following hypothesis: $2\pi v = V_{TA} k$ and $v \propto k^{\frac{D_f}{d_f}}$ are continuous, where $V_{TA}$ is the sound velocity of the TA mode. When we define $\hat{v} = \frac{v}{v_{BP}}$, $\hat{k} = \frac{k}{k_{BP}}$, $2\pi v_{BP} = V_{TA} k_{BP}$, we therefore obtain the dispersion relation as:

$$\begin{cases} \hat{v} = \hat{k}, & k < k_{BP} \\ \hat{v} = \hat{k}^{2\frac{D_f}{d_f}}, & k > k_{BP} \end{cases}, \quad (17)$$

and the coupling coefficient as:

$$C_{IR}(v) = \begin{cases} A + B v_{BP}^2 \hat{v}^2, & v < v_{BP} \\ A + B v_{BP}^2 \hat{v}^{2\frac{d_f}{D_f}}, & v > v_{BP} \end{cases}. \quad (18)$$

This will naturally facilitate continuity of $C_{IR}(v)$. Here, at $k = k_{BP}$, it is necessary that $k$ is not differentiable but continuous. We specify that the contribution of the LA mode is neglected in this hypothesis. Based on this hypothesis for the dispersion relation, we now discuss estimation of the correlation length for the BP mode and characteristic lengths for the end of the fracton.

Fig. 5a shows the dispersion relations described by the equations:

$$2\pi\nu = V_{\text{TA}}\, k, \tag{19}$$

and

$$2\pi\nu = V_{\text{TA}} k_{\text{BP}}^{1-\frac{D_f}{d_f}} k^{\frac{D_f}{d_f}}. \tag{20}$$

In the dispersion relation, we use the sound velocities of the TA mode, $V_{\text{TA}}$, and the LA mode, $V_{\text{LA}}$, of lysozyme as $V_{TA} = 1669$ m/s and $V_{LA} = 3305$ m/s at 300 K, determined through Brillouin measurements in a study by Perticaroli et al. [55]. As the fracton dispersion relation indicated by Eq. (20) has an exponent of $D_f/d_f > 1$, the dispersion relation shows an exponential function with an upward curve (with a larger monotone increasing rate) compared with the dispersion relation of the TA mode. At the BP frequency, owing to the end frequency of the dispersion relation of the TA mode, the correlation length for the BP of lysozyme, $\xi_{\text{BP}}$, is estimated to be 2.8 nm according to the relation $\xi_{\text{BP}} \approx V_{\text{TA}}/\nu_{\text{BP}}$, which represents the size distribution of the lowest-energy vibrational mode localised in a blob[63]. This correlation length, $\xi_{\text{BP}} = 2.8$ nm (see Figs. 5a and 5c), is smaller than the hydrodynamic diameter of a single lysozyme molecule, $d_{\text{LYS}} = 3.72$ nm[62] and larger than the average size of an amino acid residue (approximately 0.4 nm)[64]. This correlation length indicates a distance that contains approximately 7 amino acids and corresponds to the volume of a sphere containing 180 amino acids within radius $R = \xi_{\text{BP}}$ from the centre of gravity.

Now, we discuss the origin of the BP observed in the protein system. Regarding the correlation length, the fact that $\xi_{\text{BP}} < d_{\text{LYS}}$ apparently suggests that the origin of the BP in the protein is owing to the intrinsic disorder inside a single protein molecule rather than the disordered arrangement of the protein molecules with respect to each other. In addition, we point out that the relevance between the structure and BP of proteins is similar to that of polymer glasses. In general, polymer glass is expressed as a substance with a chain structure produced through the covalent bonding of monomer molecules. Recent MD simulations have shown that the BP frequencies of monomer and polymer glasses do not differ fundamentally; they are determined by the non-covalent bonds that exist between all monomers and not by the covalent bonds that connect the monomers[65,66]. In proteins, there are several kinds of non-covalent bonds with different magnitudes of force, such as weak van der Waals forces and hydrogen bonds, and the determinants of BP frequency are not evident. A quantitative understanding of the determinants of BP frequency in protein systems remains an open problem.

Finally, we discuss the end mode of the fracton using the dispersion relation above the BP frequency. Recent MD simulation studies have shown v-DOS behaviour in the low-frequency range of polymer glasses[65,66]. In the study by Milkus et al., the broad peak of the v-DOS from low-frequency non-covalent vibration modes is referred to as the Lennard-Jones (LJ) sea[65]. These vibrations are classified and analysed in the direction of an along-chain, a perpendicular component, and out-of-plane rocking motion with respect to a polymer chain. Then, in the LJ sea, the v-DOS peak of the along-chain motion is shown to appear at a lower frequency than that of the perpendicular motion. Here, we attempt to interpret our experimental results based on the calculation results. As a simple speculation, the peaks of the along-chain motion and perpendicular motion bands shown in the MD calculation are considered to correspond to the peak of 2.53 THz in the Raman active $\chi''(v)$ and the peak of 3.26 THz in the IR active $\varepsilon''(v)$, respectively. Furthermore, the wavelength of the mode at each peak frequency of $\varepsilon''(v)$ and $\chi''(v)$ can be estimated by considering the dispersion relation of the fracton (see Figs. 5a and 5b). The size of the end mode of the fracton in the Raman spectrum is 1.32 nm, and the value corresponds to a distance containing approximately 3.3 amino acids. In contrast, the IR active mode has a size of 1.16 nm, which corresponds to a distance containing 2.9 amino acids. Finally, we discuss the end mode of the fracton in the v-DOS spectrum. Regarding the result of the v-DOS spectrum observed by Lushnikov *et al*., the end of the fracton region is around 7.1 THz[67]. We find the size of the end mode to be 0.79 nm, and this length corresponds to the size of approximately 2 amino acids. From the above-mentioned discussion on the size of the fracton mode, we conclude that the vibration of the fracton region is manifested by the self-similar structure of the amino acid molecule and the non-covalent vibrational modes. In future, the validity of these considerations should be clarified through intensive studies on molecular dynamics calculations for proteins.

## Discussion

This study has shown that universal excitations in disordered systems, such as the BP, and the fractons of fractal objects are detectable by THz spectroscopy. For detection of fractons, we formalised the IR light-vibration coupling coefficient, $C_{IR}(v)$, in the fracton region by combining the $C_{IR}(v)$ model for disordered systems proposed by Taraskin *et al*.[47] and the fracton theory proposed by Alexander and Orbach[4]. We found that the $C_{IR}(v)$ and $C_{Raman}(v)$ in the fracton region are almost identical and exhibit the exponent of $2d_f/D_f$. To verify $C_{IR}(v)$, we performed THz time-domain and low-frequency Raman spectroscopies on the protein lysozyme that essentially exhibits a disordered and self-similar nature in a single supramolecule. The exponents of coupling coefficients were successfully explained using the values of fractal and fracton dimensions. By assuming the continuity of the dispersion relation between the TA mode and the fracton, we concluded that the fracton originates from the self-similarity of the structure of the amino acids of a single protein molecule. The fact that THz light can capture the BP and fracton will be key to understanding these universal dynamics in the nanoscale regions of disordered systems.

## Materials and Methods

**Sample preparation.** Lyophilised lysozyme powder from hen egg white at room temperature under atmospheric pressure was used, without further purification. This is a representative global type of single-domain protein with 129 amino acid residues and a molecular weight of 14.3 kDa. The sample used for this study was purchased from Wako Pure Chemical Industries, Ltd. For measurement, a pressure of 1 MPa was applied to the lysozyme powder to prepare a disk-shaped pellet.

**Terahertz time-domain spectroscopy.** THz-TDS (RT-10000, Tochigi Nikon Co.) and a liquid-helium flow cryostat system (Helitran LT-3B, Advanced Research Systems)[28,30] were used with the standard transmission configuration for temperature-dependent measurements in the frequency range of 0.25–2.25 THz from 13 to 295 K. Low-temperature grown GaAs photoconductive antennae were used as emitters and detectors of the THz wave. The THz wave propagation path was enclosed in a dry air chamber within which dry air flows. Another THz-TDS (TAS7500SU, Advantest Corp.)[28] was used as room-temperature broadband transmission THz spectroscopy. Cherenkov radiation from a lithium niobate single crystal was used as the light source of the THz wave, covering a frequency band of around 0.5–7.0 THz. Asynchronous optical sampling technology was used for the detection. To obtain a sufficient signal-to-noise ratio in every cycle, 16384 measurements were performed. The THz wave propagation path was purged with dry air.

**Low-frequency Raman scattering.** Confocal micro-Raman measurements were performed with a depolarised backscattering geometry[68]. A frequency-doubled diode-pumped solid-state Nd:yttrium-aluminum-garnet laser oscillating in a single longitudinal mode at 532 nm (Oxxius LMX-300S) was employed as the excitation source. A homemade microscope with ultranarrow-band notch filters (OptiGrate) was used to focus the excitation laser and collect the Raman-scattered light. The scattered light was analysed using a single monochromator (Jovin-Yvon, HR320, 1200 grooves/mm) equipped with a charge-coupled device camera (Andor, DU420).

**MEM analysis of specific heat data.** Under the approximation of harmonic vibration, the lattice specific heat corresponds to the v-DOS of a solid, $g(v)$, according to the following equation:

$$C(T) = 3nk_\text{B} \int \frac{x^2 e^x}{(e^x-1)^2} g(v) dv, \qquad (21)$$

where $n$ is the number of atoms and $x = hv/k_\text{B}T$. An inversion problem to obtain $g(v)$ from the specific heat data was solved using the MEM[69,70]. The MEM was performed on the specific heat data over the temperature range of 1.7–23 K[25]. The error in specific heat data was assumed to be 2%.

## Footnotes

Author contributions: T.M. conceived this study and formalised the IR light-vibration coupling coefficient in the fracton region. Y.J. performed THz-TDS measurements and analysed data. Y.F. and A.K. implemented the low-frequency Raman scattering system, performed measurements, and analysed data. H.T. and Se. K. implemented THz-TDS equipment. Y.J., K.S., and Y.Y. prepared lysozyme samples. Su. K. performed MEM analysis for low-temperature specific heat data. L.M. calculated the mass fractal dimension of the lysozyme molecule. T.M., Y.F., and H.M. discussed the results. T.M. and Y.J. wrote the paper, with contributions from and discussions with all authors.

[1]To whom correspondence should be addressed. e-mail: mori@ims.tuskuba.ac.jp

## Acknowledgments

We thank Kang Kim and Atsushi Ikeda for useful discussions. T.M. thanks Shin-ichi Kimura for the construction of the automatic measurement THz-TDS system. This work was supported by JSPS KAKENHI Grant Nos. 17K14318, 18H04476, 17K18765, 19K14670, and 16H02081, the Nippon Sheet Glass Foundation for Materials Science and Engineering, and the Asahi Glass Foundation.

## Figure Legends

**Fig. 1. THz and Raman spectra of lysozyme.** (a) Real $\varepsilon'(v)$ and (b) imaginary $\varepsilon''(v)$ parts of complex dielectric constant, and (c) BP plot $\alpha(v)/v^2$ of lysozyme determined by THz-TDS measurements at room temperature. Temperature dependence of IR spectra are shown in the insets of (a), (b) and (c), as measured in a heating process with a temperature range of 13 K to 295 K and plotted every 10 K above 20 K. (d) Measured Raman intensity $I(v)$ of lysozyme at room temperature. (e) Imaginary part of Raman susceptibility $\chi''(v)$ and (f) BP plot of Raman spectra $\chi''(v)/v$. For convenience, 1 THz = 33.3 cm$^{-1}$ = 4.14 meV = 48 K.

**Fig. 2. Comparison of THz, Raman, and v-DOS spectra.** (a) Imaginary parts of complex dielectric constant $\varepsilon''(v)$ (orange) and Raman susceptibility $\chi''(v)$ (blue) of lysozyme. (b) BP plots of IR ($\alpha(v)/v^2$ (red)) and Raman ($\chi''(v)/v$ (green)) spectra. (c) BP plot of v-DOS $g(v)/v^2$. The data of $g(v)$ are quoted from previous studies based on INS[40,55]. The fitting result obtained with a log-normal function[56] to depict the BP frequency is shown by the pink line.

**Fig. 3. Log-log plot of THz, Raman, and v-DOS spectra and coupling coefficients.** (a) Log-log plot of imaginary parts of complex dielectric constant $\varepsilon''(v)$ (orange), Raman susceptibility $\chi''(v)$ (arb. units) (blue), and v-DOS $g(v)$ (cyan) of lysozyme. $g(v)$ data are quoted from previous studies based on INS[40,55]. (b) IR light-vibration (red) and Raman (green) coupling coefficients ($C_{IR}(v)$ and $C_{Raman}(v)$) of lysozyme in the log-log representation. The inset shows the linear representation of $C_{IR}(v)$ and $C_{Raman}(v)$ in the vicinity of the BP frequency.

**Fig. 4. Calculation of mass fractal dimension of lysozyme molecule.** (a) Atoms (red balls) contained in a sphere of radius $R$ from the centre of gravity of the lysozyme molecule. For the lysozyme atomic coordinate arrangement, information on 1LYZ was taken from the PDB. (b) Log-log plot of the $R$ dependence of molecular weight contained in a sphere of radius $R$ from the centre of gravity of lysozyme. When the diameter of the spheres exceeds the diameter of the lysozyme molecules, the molecular weight saturates. From fitting to the linear region, the mass fractal dimension of the lysozyme molecule was obtained as $D_{f\text{-calc}} = 2.75$.

**Fig. 5. Dispersion relation in fracton region and THz and Raman spectra.** (a) Dispersion relation of TA and fracton modes of lysozyme with the hypothesis of continuous connection at the BP frequency (green solid line). The LA (red dashed line) and TA (grey dashed line) modes are also shown, with sound velocity as measured by Perticaroli *et al.*[55] The dotted lines represent the structure correlation length (wine red dashed line) and characteristic wavelengths (blue (Raman) and orange (IR) dashed lines). These lengths are indicated by bars

on the figure of the lysozyme molecule. (b) Imaginary parts of complex dielectric constant $\varepsilon''(v)$ (orange) and Raman susceptibility $\chi''(v)$ (blue) of lysozyme. (c) BP plots of IR ($\alpha(v)/v^2$ (red)) and Raman ($\chi''(v)/v$ (green)) spectra.

## Table Legend
**Table 1. Exponent values of the coupling coefficients in the fracton region, independently obtained fractal and fracton dimensions, and $2d_f/D_f$.**

(Left) Exponent values of $C_{IR}(v)$ and $C_{Raman}(v)$ in the fracton region. Fitting lines are shown in Fig. 3b. (Right) Numerically ($D_{f\,calc}$ and $d_{f\,calc}$) and experimentally ($D_{f\,exp}$ and $d_{f\,exp}$) obtained fractal and fracton dimensions, and calculated $2d_f/D_f$. A fitting line to extract $D_{f\text{-}calc}$ is shown in Fig. 4b.

**Fig. 1.**

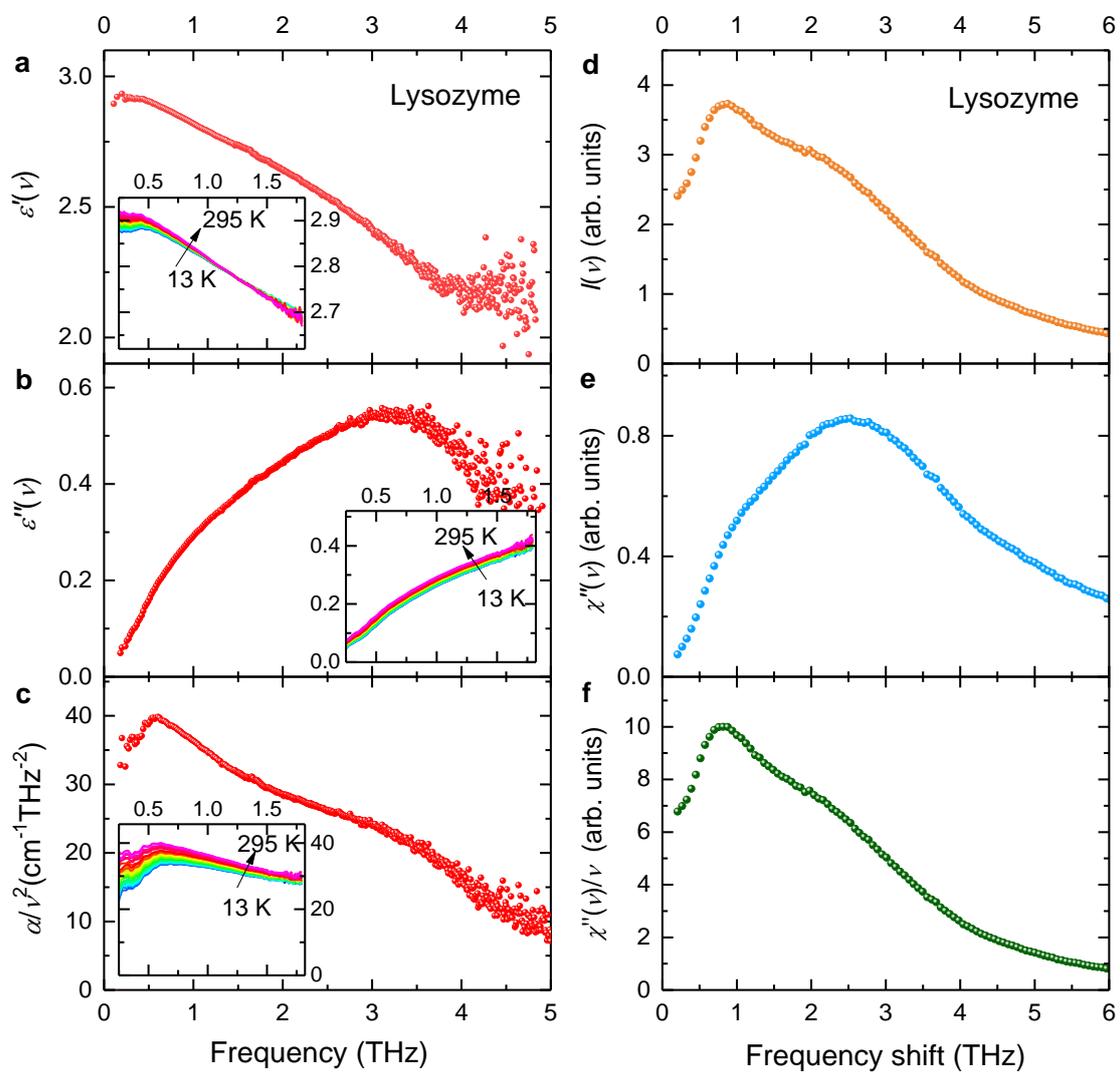

**Fig. 2.**

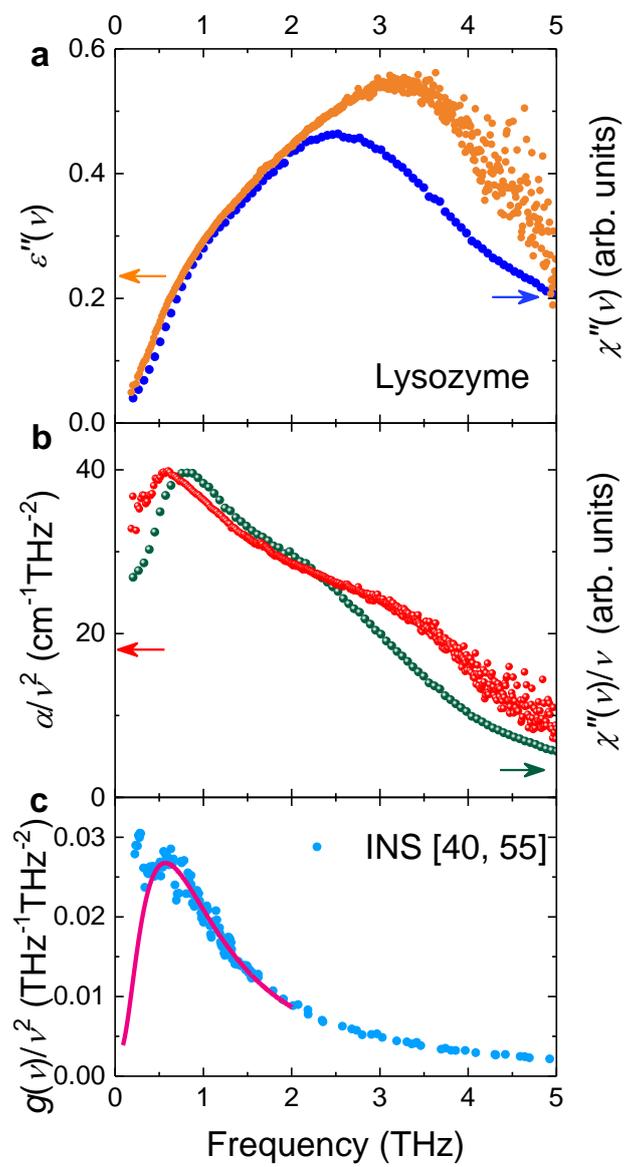

**Fig. 3.**

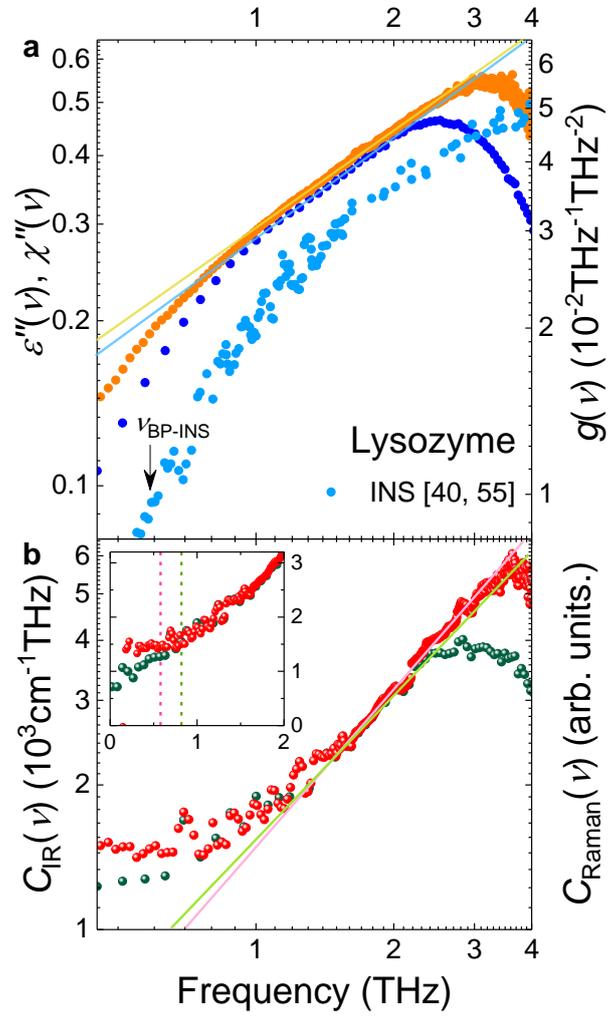

**Fig. 4.**

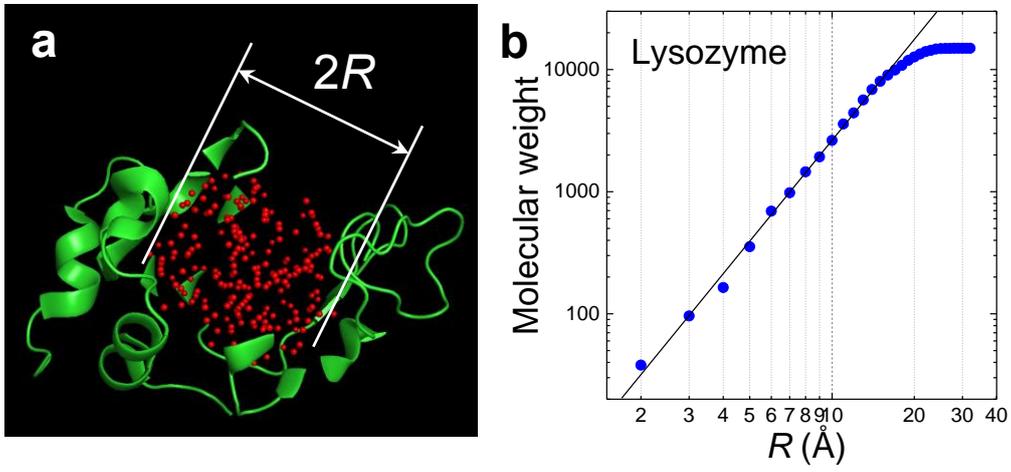

**Fig. 5.**

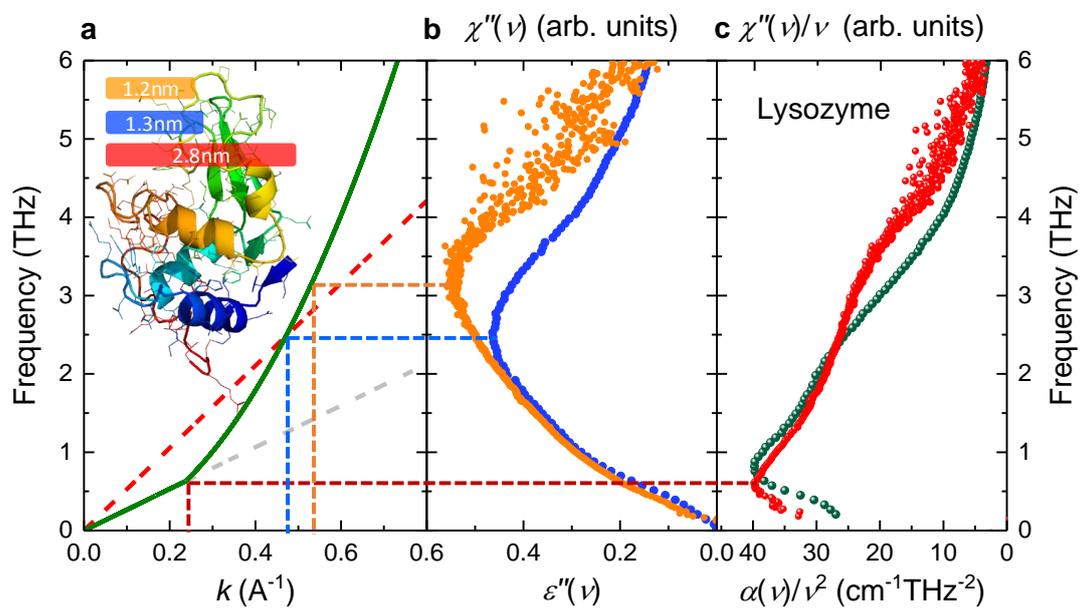

**Table 1.**

| | Exponent value of $C_{IR}(v)$ and $C_{Raman}(v)$ | | $D_f$ | $d_f$ | $2\dfrac{d_f}{D_f}$ |
|---|---|---|---|---|---|
| IR | 1.10 | Calculation | 2.75 | 1.43[61] | 1.04 |
| Raman | 1.00 | Experiment | 2.78[40] | 1.43[40] | 1.02 |

# Supplementary Information

# Detection of boson peak and fractal dynamics of disordered system using terahertz spectroscopy

Tatsuya Mori[a,1], Yue Jiang[a], Yasuhiro Fujii[b], Suguru Kitani[c], Hideyuki Mizuno[d], Akitoshi Koreeda[b], Leona Motoji[a], Hiroko Tokoro[a], Kentaro Shiraki[a], Yohei Yamamoto[a], Seiji Kojima[a]

[a]Division of Materials Science, University of Tsukuba, 1-1-1 Tennodai, Tsukuba, Ibaraki 305-8573, Japan

[b]Department of Physical Sciences, Ritsumeikan University, 1-1-1 Nojihigashi, Kusatsu 525-8577, Japan

[c]Laboratory for Materials and Structures, Tokyo Institute of Technology, 4259 Nagatsuta-cho, Midori-ku, Yokohama 226-8503, Japan

[d]Graduate School of Arts and Sciences, The University of Tokyo, Tokyo 153-8902, Japan

1Corresponding author
Email: mori@ims.tuskuba.ac.jp

**Table S1.** Summary of BP plot relations. The rightmost parts of Eq. S2 represent the high-temperature approximation of Raman spectra.

$$\alpha(\nu) = C_{\text{IR}}(\nu)g(\nu) \tag{S1a}$$

$$\frac{\alpha(\nu)}{\nu} = C_{\text{IR}}(\nu)\frac{g(\nu)}{\nu} = \frac{2\pi\mu'}{cn(\nu)}\varepsilon''(\nu) = \frac{4\pi}{c}\kappa(\nu) \tag{S1b}$$

$$\frac{\alpha(\nu)}{\nu^2} = C_{\text{IR}}(\nu)\frac{g(\nu)}{\nu^2} \tag{S1c}$$

$$\nu\chi''(\nu) = C_{\text{Raman}}(\nu)g(\nu) = \frac{\nu I(\nu)}{n_{\text{B}}(\nu,T)+1} \approx \nu^2 I(\nu) \tag{S2a}$$

$$\chi''(\nu) = C_{\text{Raman}}(\nu)\frac{g(\nu)}{\nu} = \frac{I(\nu)}{n_{\text{B}}(\nu,T)+1} \approx \nu I(\nu) \tag{S2b}$$

$$\frac{\chi''(\nu)}{\nu} = C_{\text{Raman}}(\nu)\frac{g(\nu)}{\nu^2} = \frac{I(\nu)}{\nu[n_{\text{B}}(\nu,T)+1]} \approx I(\nu) \tag{S2c}$$

where,

$\alpha(\nu)$ absorption coefficient

$C_{\text{IR}}(\nu)$ infrared light-vibration coupling coefficient

$g(\nu)$ vibrational density of states

$\mu'$ real part of permeability

$c$ speed of light

$n(\nu)$ refractive index

$\varepsilon''(\nu)$ imaginary part of complex dielectric constant

$\kappa(\nu)$ extinction coefficient

$\chi''(\nu)$ imaginary part of Raman susceptibility

$C_{\text{Raman}}(\nu)$ Raman coupling coefficient

$I(\nu)$ measured Raman intensity

$n_{\text{B}}(\nu,T) = (\exp(h\nu/k_B T) - 1)^{-1}$ Bose-Einstein distribution function

**Table S2** (Top) Parameters used for the calculation of uncorrelated charge $q_1$ in $C_{IR}(v)$ for the lysozyme. Experimentally determined constant term $A$, average mass $\bar{m}$, density $\rho$, dielectric constant at 0.8 THz instead of the high-frequency dielectric constant $\varepsilon_\infty$, and obtained uncorrelated charge $q_1$ for the lysozyme. (Bottom) Comparison between lysozyme and silica glass. The data of the silica glass were measured separately and the values are consistent with a previous report[1,2].

|  | $A$(cm$^{-1}$THz) | $\bar{m}$ (g) | $\rho$ (g cm$^{-3}$) | $\varepsilon_\infty$ | $q_1(e)$ |
|---|---|---|---|---|---|
| Lysozyme | 1399 | $1.28 \times 10^{-23}$ | 0.869 | 2.19 | 0.127 |

|  | Lysozyme | Silica glass | Comparison with silica (Lyz/Silica) |
|---|---|---|---|
| $\alpha$ (cm$^{-1}$) | 25 (at 0.8 THz) | 1.75 (at 1 THz) | 14.29 |
| $A$ (cm$^{-1}$THz) | 1399 | 899 | 15.57 |
| $q_1(e)$ | 0.127 | 0.06 | 2.12 |
| $\langle q_{1i}^2 \rangle$ |  |  | 4.48 |
| $\bar{m}$ ($10^{-23}$g) | 1.28 | 3.37 | 0.38 |
| $1/\bar{m}^2$ |  |  | 6.93 |
| $\rho$ (g cm$^{-3}$) | 0.869 | 2.21 | 0.39 |
| $\varepsilon_\infty$ | 2.19 | 3.81 | 0.57 |
| $\dfrac{2\pi^2 \rho}{\bar{m}^2 c \sqrt{\varepsilon_\infty}}$ |  |  | 3.59 |

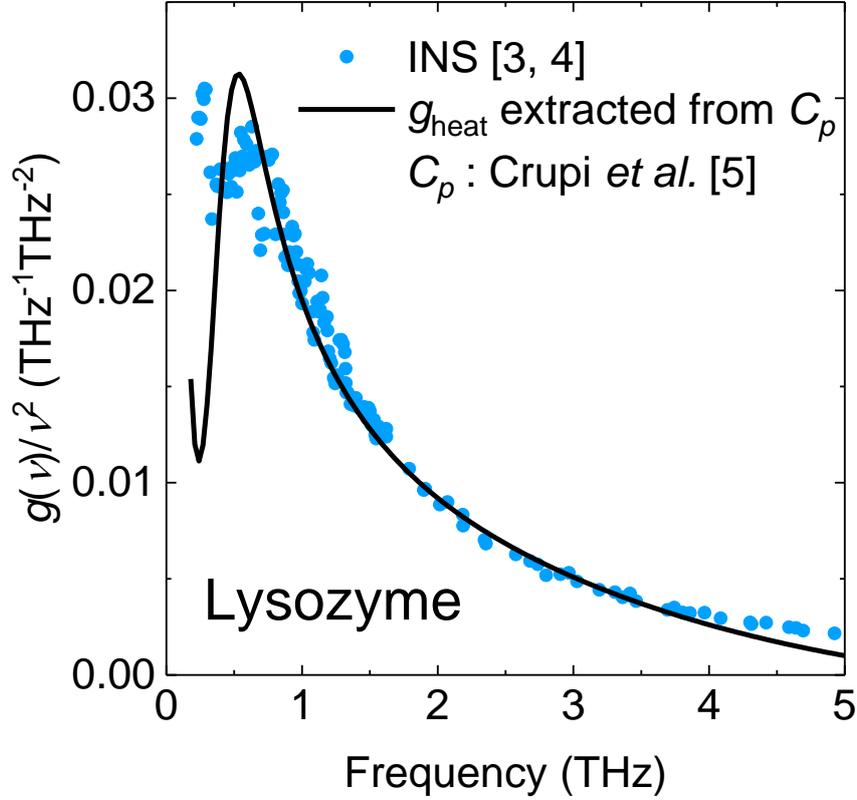

Fig. S1. Adjustment of $g(v)$ by inelastic neutron scattering with the absolute value determined by performing MEM analysis on the low-temperature specific heat ($C_p$) data of lysozyme. We depicted and connected $g(v)$ from inelastic neutron scattering measurements by Perticaroli et al.[3] and Lushnikov et al.[4] for achieving higher accuracy of $g(v)$[3] in the BP region and $g(v)$[4] in the fracton region. Based on data depicted from the results of low-temperature specific heat measurements reported in the work by Crupi et al.[5], the specific heat v-DOS as $g_{heat}(v)$ was extracted by MEM analysis. We determined the absolute value of the $g_{INS}(v)$ used in this work through conformation with the absolute value of $g_{heat}(v)$ at the BP frequency.